\documentclass[
    ,final            
  ]
  {aipproc}

\layoutstyle{8x11single}

\newcommand{\perkmd}{\ensuremath{\textrm{keV}^{-1}\textrm{m}^{-2}\textrm{day}^{-1}}} 

\begin{document}

\title{The BetaCage, an ultra-sensitive screener for surface contamination}


\author{R.~Bunker}{
  address={Department of Physics, Syracuse University, Syracuse, NY 13244, USA}
}

\author{Z.~Ahmed}{
	address={California Institute of Technology, Pasadena, CA 91125, USA}
}

\author{M.A.~Bowles}{
  address={Department of Physics, Syracuse University, Syracuse, NY 13244, USA}
}

\author{S.R.~Golwala}{
	address={California Institute of Technology, Pasadena, CA 91125, USA}
}

\author{D.R.~Grant}{
	address={University of Alberta, Edmonton, AB, T6G 2R3, Canada}
}	

\author{M.~Kos}{
  address={Department of Physics, Syracuse University, Syracuse, NY 13244, USA},
  altaddress={Pacific Northwest National Laboratory, Richland, WA 99352, USA} 
}

\author{R.H.~Nelson}{
	address={California Institute of Technology, Pasadena, CA 91125, USA}
}

\author{R.W.~Schnee}{
  address={Department of Physics, Syracuse University, Syracuse, NY 13244, USA}
}

\author{A.~Rider}{
	address={California Institute of Technology, Pasadena, CA 91125, USA}
}

\author{B.~Wang}{
  address={Department of Physics, Syracuse University, Syracuse, NY 13244, USA}
}

\author{A.~Zahn}{
	address={California Institute of Technology, Pasadena, CA 91125, USA}
}

\begin{abstract}
Material screening for identifying low-energy electron emitters and alpha-decaying isotopes is now a prerequisite for rare-event searches (\itshape e.g.\normalfont, dark-matter direct detection and neutrinoless double-beta decay) for which surface radiocontamination has become an increasingly important background. The BetaCage, a gaseous neon time-projection chamber, is a proposed ultra-sensitive (and nondestructive) screener for alpha- and beta-emitting surface contaminants to which existing screening facilities are insufficiently sensitive. 
Sensitivity goals are 0.1 betas keV$^{-1}$\,m$^{-2}$\,day$^{-1}$ and 0.1 alphas~m$^{-2}$\,day$^{-1}$, with the former limited by Compton scattering of photons in the screening samples and (thanks to tracking) the latter expected to be signal-limited; radioassays and simulations indicate backgrounds from detector materials and radon daughters should be subdominant. We report on details of the background simulations and detector design that provide the discrimination, shielding, and radiopurity necessary to reach our sensitivity goals for a chamber with a $95\times95$\,cm$^{2}$ sample area positioned below a 40\,cm drift region and monitored by crisscrossed anode and cathode planes consisting of 151 wires each.
\end{abstract}

\maketitle


\section{Introduction and detector design}
\label{sec:intro}

Nonpenetrating radiation on detector surfaces, particularly from radon daughters, provides a dominant background for many underground physics experiments.
Radon daughters from the atmosphere deposit onto surfaces and decay to the long-lived $^{210}$Pb, a low-energy beta emitter, and then to the alpha-emitting $^{210}$Po. 
Unfortunately, high-sensitivity detection of the $^{210}$Pb 46\,keV gamma ray from material surfaces is generally not feasible with HPGe detectors, in part due to its low branching fraction.
Furthermore, since these surface contaminants are chemically separated and hence out of equilibrium with the
photon-emitting isotopes in the parent $^{238}$U decay chain,
direct detection of
the $^{210}$Po alphas or betas from $^{210}$Pb or $^{210}$Bi
is necessary to establish the surface contamination level.
More sensitive detection of alphas and betas on surfaces would also serve 
archeology, biology, climatology, environmental science,
geology, integrated-circuit quality control, and planetary science~\cite{schneeLRT2006}.

The BetaCage~\cite{schneeLRT2006,shutt_lrt2004,ahmedLRT2010},  an ultra-low-background drift chamber, 
provides the largest efficiency possible for detection of alphas and $<200$\,keV electrons
from a sample's surface by eliminating backscattering and dead-layer
effects while providing a large sensitive area.
The detector uses
the minimum amount of gas needed to stop particles of interest in
order to minimize background from ambient penetrating gammas.  
It has the minimum possible surface area that itself
can be a source of background particles.  Finally, it
provides sufficient spatial information to distinguish
events coming from the sample surface from those due to
scattering of background particles in the
gas.

\begin{figure}[tb!]
\centering
\includegraphics[height=2.0in] {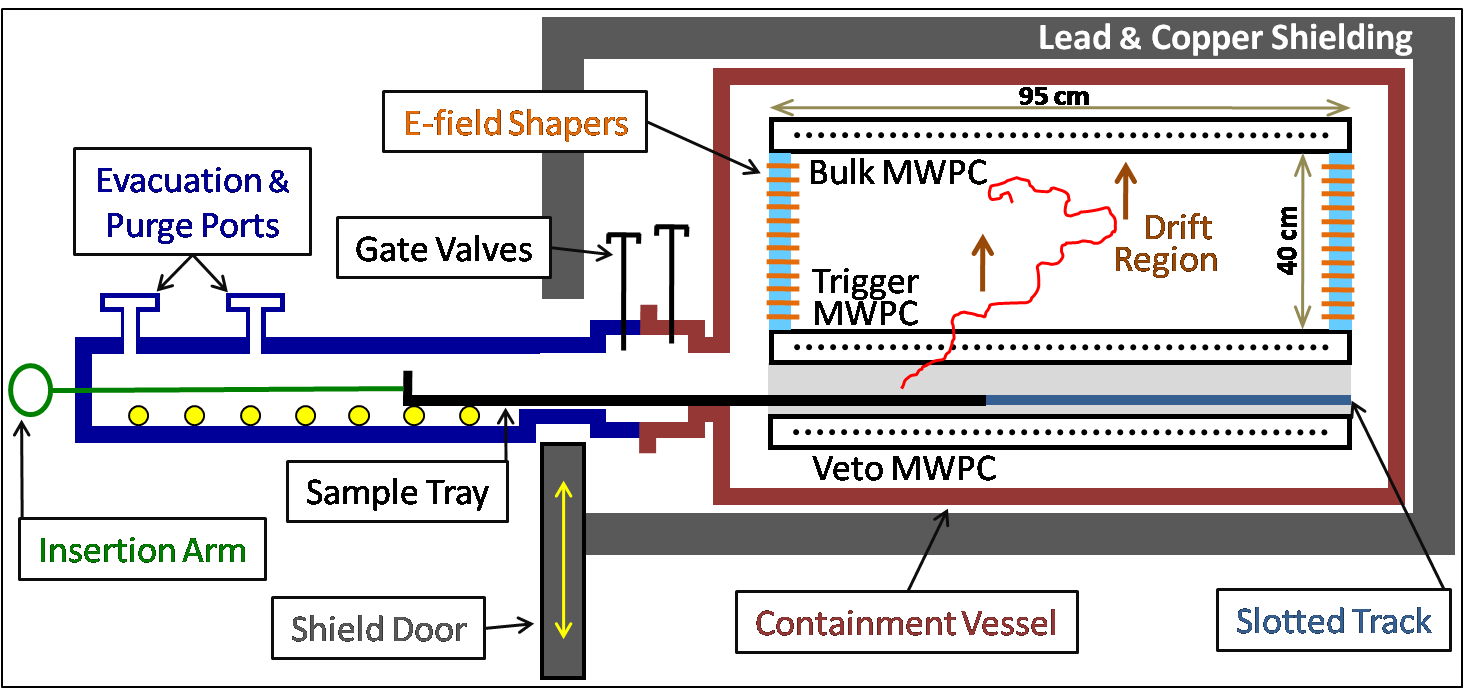}
\includegraphics[height=2.0in] {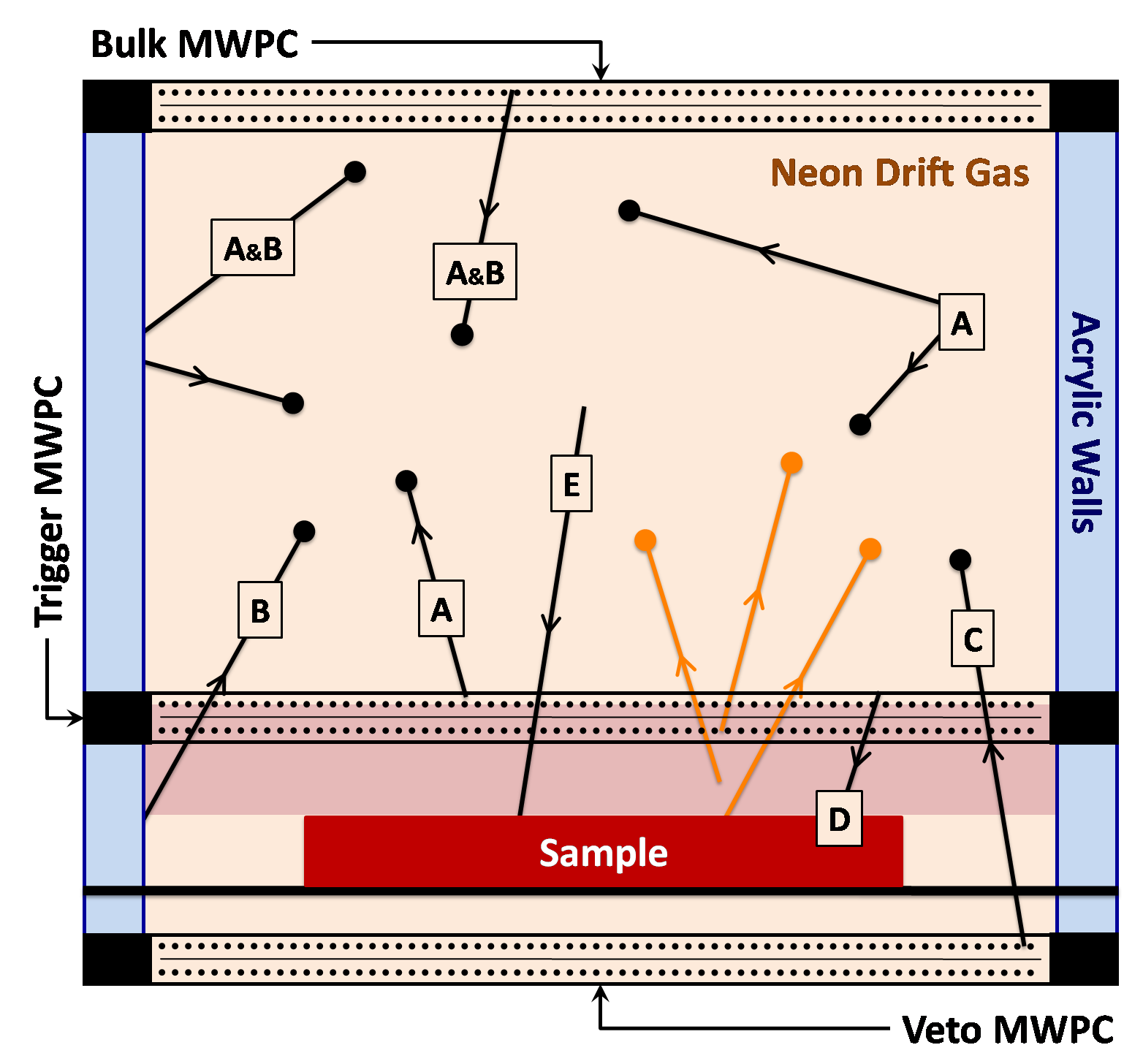} 
\caption{Schematic side views of the BetaCage.  
The sample lies between the $95\times95$\,cm$^{2}$ veto and trigger MWPCs, 40\,cm below the bulk MWPC.
 {\it Left}: The
   shielding consists of 15\,cm lead outside
 5\,cm low-activity 
 lead and a
 1-cm thick OFE Cu liner, surrounding a PMMA 
 vessel.  Not shown is a  2-mm thick stainless-steel dogleg to feedthrough signals. 
  {\it Right}: 
  Examples of background tracks rejectable (dark lines) due to their (A) insufficient energy in the trigger MWPC, (B) lack of containment in the fiducial drift region, (C) too much energy in the veto MWPC, (D) insufficient energy in the bulk MWPC, or (E) 
  uniform $dE/dx$, even at the apparent track end.
  Rejection is less effective (light lines) for betas or alphas emitted from the drift gas and cathode wires directly above the sample, or from the surface of the sample itself, following Rn-daughter implantation or Compton scattering.
}
\label{fig:betacage}
\end{figure}

Figure~\ref{fig:betacage} shows sketches of the proposed BetaCage.
Samples are placed in the gas directly under an open $95\times95$\,cm$^{2}$ 
multi-wire proportional counter (MWPC) that
provides a trigger for particles emanating from the sample.
Above this ``trigger'' MWPC is a 40-cm-high region, large enough to contain the full tracks of alphas and 200\,keV betas. 
An electric field in this region drifts the ionization
to the top of the chamber, where a second open (``bulk'') MWPC
collects it.  
Proportional avalanching and crossed grids in both MWPCs provide gain and $xy$-position determination, respectively.
The time profile of charge collection in the bulk MWPC determines the
spatial profile of the track in the $z$ dimension, while the 
trigger MWPC provides a start time to establish absolute $z$ location.
The right panel of Fig.~\ref{fig:betacage} demonstrates how the BetaCage achieves excellent rejection of background events, especially when used for alpha screening (since alpha tracks are straight enough not to suffer reconstruction failures).

\section{Expected photon-induced backgrounds}

\begin{figure}[tb!]
\centering
\includegraphics[height=2.3in] {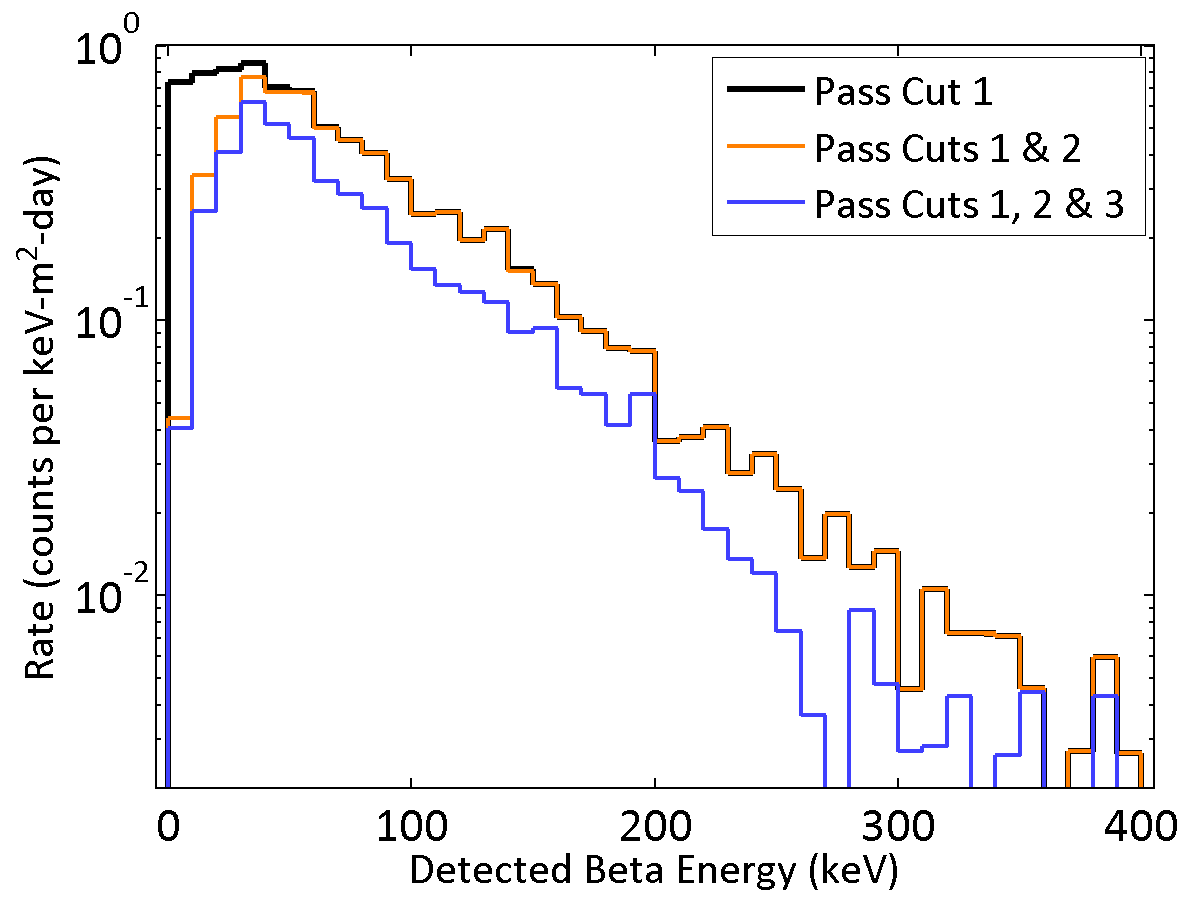} 
\includegraphics[height=2.3in] {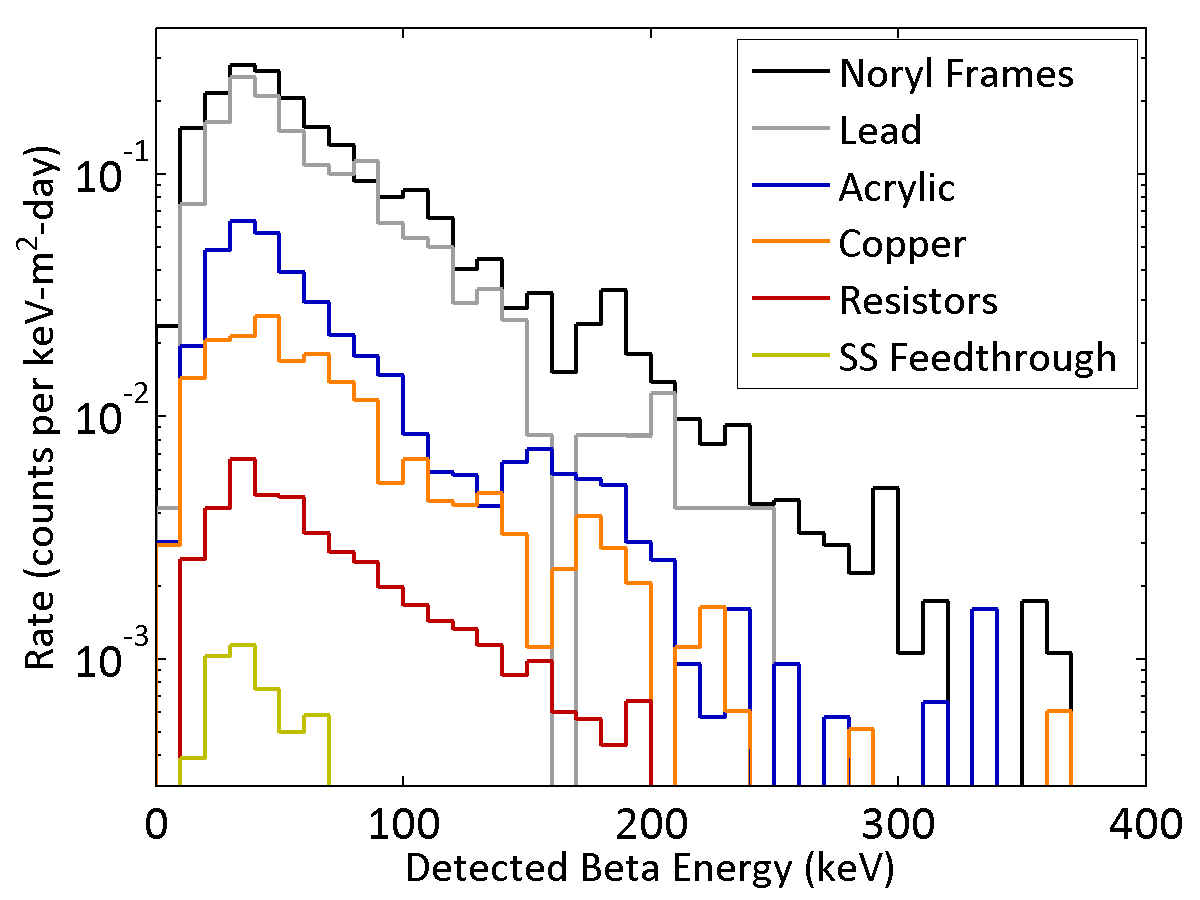} 
\caption{
{\it Left}: Expected rate of photon-induced backgrounds after cumulative data-selection cuts requiring, from top to bottom,  (1, black)
$<1$\,keV ($>1$\,keV) in the veto (trigger) MWPC,
(2) $>5$\,keV deposited in the drift region (bulk MWPC), and (3) the event be fully contained in the drift region.
{\it Right}: Division by source of expected photon-induced backgrounds after all cuts. From top to bottom, spectra are from 
contamination in the Noryl MWPC frames, lead shielding, acrylic chamber, copper liner, field-shaping resistors, and stainless-steel feedthrough.  The Noryl spectrum is based on screening upper limits and is likely conservative; the lead is expected to dominate.  Other assumed material-contamination levels are based on upper limits or measurements in~\cite{LRT2013loach,xenonEMbackgrounds2011,LRT2013resistors}.}
\label{fig:photons}
\end{figure}

The planned 
20-cm thick lead shield
should reduce backgrounds due to external photons to a sub-dominant level,
based on past efforts and simulations~\cite{daSilvathesis,r19prd}. 
The background from $^{210}$Bi in the inner 5\,cm of lead with $3$\,Bq/kg $^{210}$Pb, simulated by the technique of~\cite{vojtyla1996NIMPBshielding}, is likely to be dominant (but is expensive to reduce).
Additional 
photons are emitted from internal acrylic and copper components, and the Noryl MWPC frames.
The most challenging requirement of the detector was to identify a plastic that 
is sufficiently radiopure yet strong enough to handle the wire tension (40\,kg in each direction) 
and can be precision machined
when properly annealed.
Screening with the UMN Gopher HPGe detector and the UC Davis neutron-activation technique, and ICP-MS measurements at Caltech demonstrated that Noryl has sufficiently low
contamination levels 
($\leq1$\,mBq/kg~$^{238}$U,
3\,mBq/kg~$^{232}$Th, and 5\,mBq/kg~$^{40}$K).
Figure~\ref{fig:photons} shows the expected background spectra from photons, broken down by cuts, and by source material after all cuts.

\section{Expected Radon-induced Backgrounds}

Radon daughter plate-out onto BetaCage wires before or during construction of the MWPCs can be a dominant background.
If the initial $^{210}$Pb surface concentration on purchased wires (currently undergoing screening) proves to be larger than the background goal, 
the wire will be electropolished~\cite{ZuzelElectropolishSteel2012} with a partially automated setup prior to wire-frame construction.
We have already demonstrated the ability to reduce contamination while removing a consistent thickness of stainless-steel wire~\cite{LRT2013schneeEP}.
To minimize deposition during construction, the MWPCs will be strung within Syracuse's low-radon cleanroom~\cite{LRT2013schneeVSA}.  The Jacobi model predicts 
the achieved radon level $<1$\,Bq/m$^{3}$
is sufficiently low to keep this background sub-dominant; tests are in progress to confirm the predicted deposition rate. 

\begin{figure}[tb!]
\centering
\includegraphics[height=2.3in] {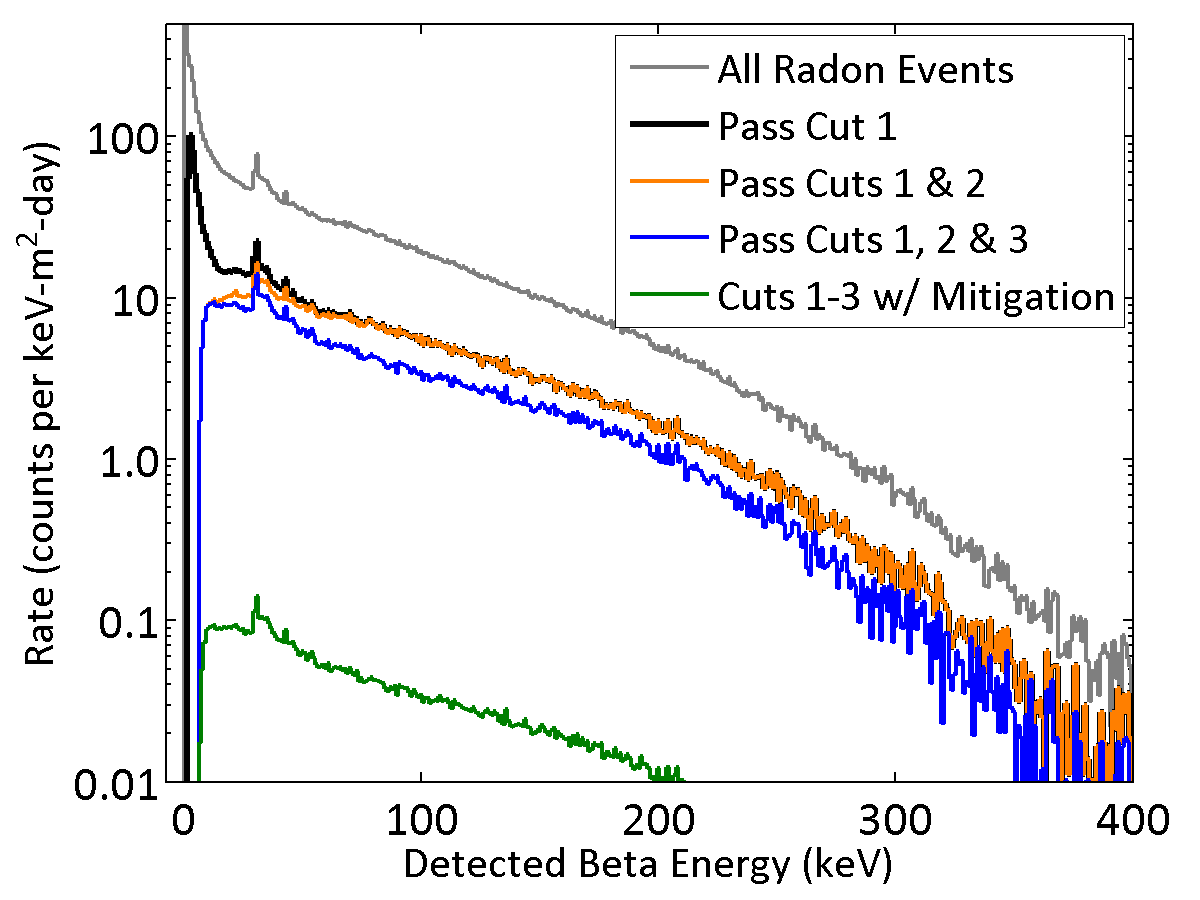} 
\includegraphics[height=2.3in] {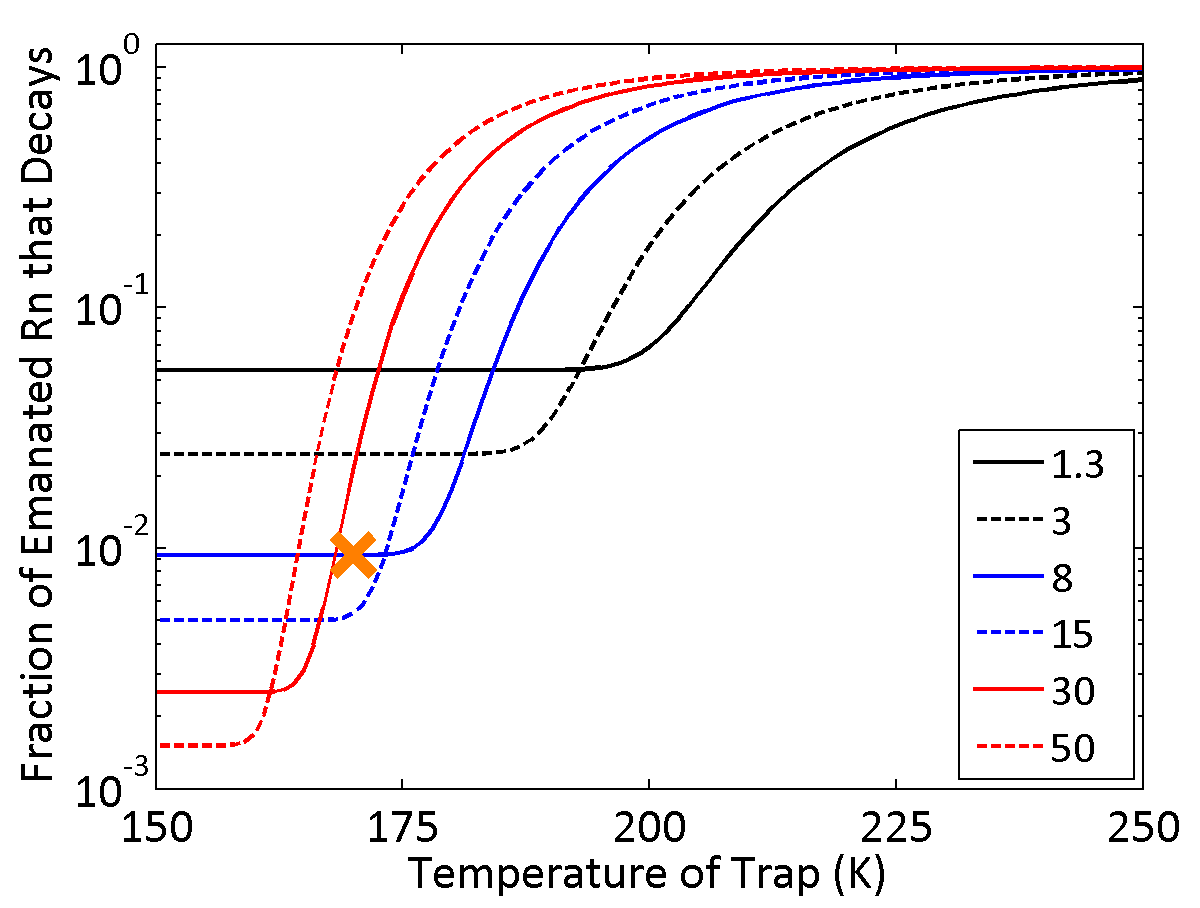} 
\caption{
{\it Left}: Expected rate of radon-induced backgrounds before and after cumulative data-selection cuts requiring, from top to bottom, (1) $<1$\,keV ($>1$\,keV) in the veto (trigger) MWPC, (2) $>5$\,keV deposited in the drift region (bulk MWPC), and (3) the event be fully contained in the drift region.
Planned Rn mitigation reduces the expected background by $\sim100\times$ (bottom spectrum).
{\it Right}: 
Fraction of emanated Rn that decays in the detection chamber rather than in the radon trap, as functions of the trap temperature, for the flow rates listed in the legend (lpm).  
The $\times$ indicates the parameters for the planned carbon trap, reducing radon-induced backgrounds by $\sim100\times$ (even if the trap's efficiency is $\sim10\times$ worse than ideal).
Below a certain temperature, 
enough Rn atoms entering the trap decay (in the trap) 
that its effectiveness is  limited by the fraction of Rn atoms that decay before they are purged from the detection chamber (via circulation), and is constant with temperature.
}
\label{fig:radon}
\end{figure}

Radon emanation from detector materials after construction similarly can be a dominant background.
The left panel of Fig.~\ref{fig:radon} shows this expected background, starting from an assumed total emanation rate supporting 1300 $^{222}$Rn\,decays/day, 
which should be readily achievable with careful materials selection.
Due to the long (22.3\,year) half-life of $^{210}$Pb, the rate of the late beta decays in the radon chain ($^{210}$Pb and $^{210}$Bi) will be lower than the rate of the early betas, by about an order of magnitude after the detector is operated for a few years.  
Here we assume 130 $^{210}$Pb and $^{210}$Bi decays/day.
Since most Rn daughters are positively charged, most will 
plate out onto the most negative nearby surface, typically the top cathode wire of the trigger MWPC.  We assume 70\% of the decays occur on these wires, 25\% on the veto MWPC wires, and 5\% in the gas volume.  Simulations indicate that data-selection cuts reduce the resulting background rate by about an order of magnitude, 
as shown in the left panel of Fig.~\ref{fig:radon}.

In order to reduce this (otherwise dominant) rate further, the gas-handling system will include a cooled-carbon radon trap, 25\,cm long with a 2.5\,cm diameter, including 70\,g  of high-quality synthetic carbon (Blucher GmbH 102688) with 1342\,m$^{2}$/g surface area.  For an 8-lpm circulation flow rate, 
 operation of the trap at 
 170\,K should result 
 in a 
sufficiently small survival probability for Rn atoms entering the trap that the radon-mitigation factor will be dominated by the flush time of the 0.6\,m$^{3}$ detector volume, yielding a reduction of $>100$, as shown in the right panel of Fig.~\ref{fig:radon}.
This reduction makes the expected backgrounds from radon (left panel of Fig.~\ref{fig:radon}) smaller than those  from photons.



The planned underground location of the BetaCage has a high 
radon level, $\sim1000$\,Bq/m$^{3}$.  A low-radon purge around the PMMA vessel will mitigate this background.  The radon level requires that the gas handling system be leak-tight to $\sim10^{-6}$\,mbar\,L/s downstream of the carbon trap.  Upstream of the carbon trap, a much higher leak rate of $\sim0.01$\,mbar\,L/s is allowed, potentially permitting the use of a relatively leaky mini-diaphragm circulation pump.  The carbon trap also 
can be used, if needed, to reduce the radon level in the neon when first filling the chamber.



\section{Expected sensitivity to betas and alphas}

The expected total beta background, dominated by the 
photon-induced background ($>80\%$ of total),
is 0.25 betas~keV$^{-1}$\,m$^{-2}$\,day$^{-1}$   
from 6~to~200\,keV.  
One day of background measurement would
establish a 49~m$^{-2}$~day$^{-1}$ background rate over 6~to~200\,keV to a precision of
7.0\,m$^{-2}$~day$^{-1}$, allowing detection at $>3\sigma$ of
a 0.1\,\perkmd\ rate
contaminant in a total of 3 days of running.
Fifteen days counting for sample and background (each) would allow
sensitivity to a 0.04\,\perkmd\ rate.
Even for a background rate $10\times$ higher than expected, 
fifteen days counting for sample and background (each) would yield
sensitivity to a 0.13\,\perkmd\ rate.


The dominant background for alpha counting is expected to be from radon daughters.
Gamma rays from natural radioactivity have too little energy to be mistaken for surface alphas, and the detector's underground, shielded environment should make
unvetoed backgrounds from cosmic rays negligible.  
Head/tail track discrimination is expected to be essentially perfect.
With the detector's clean gas, low surface area of material in the fiducial region, and tracking, 
the expected background is $>100\times$ lower than the best currently achieved, 20\,m$^{-2}$~day$^{-1}$ with the XIA UltraLo-1800~\cite{XIA2007}.  In practice, alpha sensitivity limits are likely to be signal limited;
for two weeks counting, the expected sensitivity $\sim0.1$\,counts\,m$^{-2}$ day$^{-1}$. 


\begin{theacknowledgments}
This work was supported in part by the 
National Science Foundation (Grants No.\ PHY-0855525 and PHY-0919278) and the Department of Energy HEP division.
\end{theacknowledgments}



\bibliographystyle{aipproc}   

\bibliography{schnee}

 \IfFileExists{\jobname.bbl}{}
 {\typeout{}
  \typeout{******************************************}
  \typeout{** Please run "bibtex \jobname" to obtain}
  \typeout{** the bibliography and then re-run LaTeX}
  \typeout{** twice to fix the references!}
  \typeout{******************************************}
  \typeout{}
 }

\end{document}